\documentclass{jetpl}
\twocolumn

\title{A search for Extragalactic Sources of Ultrahigh-Energy Cosmic Rays}

\rtitle{A search for Extragalactic Sources of UHECR}

\sodtitle{A search for Extragalactic Sources of Ultrahigh-Energy Cosmic Rays}

\author{A.\,A.\,Ivanov\thanks[1]{for the Yakutsk array group},}

\rauthor{A.\,A.\,Ivanov}

\sodauthor{A.\,A.\,Ivanov}

\address{Shafer Institute for Cosmophysical Research and Aeronomy, Yakutsk, 677980 Russia\\
e-mail: ivanov@ikfia.ysn.ru}

\dates{11 December 2007}{17 January 2008}

\abstract{
Possible extragalactic sources of cosmic rays at energies above $4\times 10^{19}$ eV detected with the Yakutsk array are sought. Correlation of the shower arrival directions with objects from V\'{e}ron's catalog that are located closer than 100 Mpc from the Earth confirms the observations at the Pierre Auger observatory, as well as the Greisen-Zatsepin-Kuzmin effect on ultrahigh-energy cosmic rays. The detailed analysis of the data reveals the classes of objects belonging to the active galactic nuclei that are probable sources of ultrahigh-energy cosmic rays.}
\PACS{96.40.Pq, 98.54.Cm, 98.70.Sa}

\begin{document}
\maketitle
The problem of the origin of ultrahigh-energy cosmic rays (UHECRs) remains unsolved in spite of considerable efforts of researchers. The reason is that the observed distribution of the arrival directions of cosmic rays is practically isotropic, due to the trajectories of charged particles deflected in magnetic fields in the Galaxy and beyond. Indications of a possible correlation of UHECR arrival directions with certain most active extragalactic objects have been obtained only at the highest energies $E>4\times 10^{19}$ eV (= 40 EeV), where the deflection of protons, the most probable particles of cosmic rays, is comparable or smaller than the angular resolution of detectors.

In particular, Farrar and Biermann~\cite{Farrar} found correlation of the highest energy ($E>10^{19.9}$\,eV) extensive air showers (EASs) detected to that time with compact radio quasars at high redshifts. On the contrary, Glushkov~\cite{Glushkov} pointed out the correlations with closer quasars, $z<2.5$. However, these results were not confirmed by subsequent observations.

Another type of the extragalactic objects that can generate UHECRs is presented by gamma ray bursts~\cite{Waxman}. In this case, a search for correlation is complicated, because these are short-term events with duration less than several minutes. Nevertheless, the data from the Pierre Auger observatory (PAO, 609 161 EASs) were compared~\cite{GRB} with the observations by Swift, HETE, INTEGRAL, etc. missions (284 bursts detected with an angular resolution better than 1$^0$). To reveal correlation of the arrival directions with the coordinates of the bursts within the angular range from 5$^0$ to 30$^0$, a time interval of 100 days before and after each burst is taken. Comparing the event rate of cosmic rays before and after the birst as a function of time difference, Anchordoqui~\cite{GRB} concluded that there is no frequency difference detectable.

Stanev et al.~\cite{Stanev} noted that the existing set of cosmic rays with energies $E>40$ EeV detected before 1995 tends to align along the Supergalactic plane, where the density of galaxies is relatively high. However, analysis of data from five Northern-Hemisphere arrays (114 showers, $E>40$ EeV) performed in 2000~\cite{Uchihori} revealed no significant excess of particles from the plane. At the same time, Uchihori et al.~\cite{Uchihori} noted that the number of doublets and triplets (coincidences in the arrival directions of two and three particles, respectively, within $4^0$) in $\pm10^0$ band near the Supergalactic plane is larger than that expected for the isotropic distribution. They treated it as the possibility for the part of cosmic rays at energies above $4\times10^{19}$ eV to correlate with the Supergalactic plane.

Blasars, including BL Lacertae (BL Lacs) and OVV quasars, used to be regarded as active galactic nuclei (AGNs) with relativistic jets pointing at the Earth. Tinyakov and Tkachev~\cite{TT} found a significant angular correlation of UHECRs detected by the AGASA (39 EASs, $E>48$ EeV) and Yakutsk (26 EASs, $E>24$ EeV) arrays with BL Lacs ($m < 18$) from catalog~\cite{VCV}. The analysis of HiRes data performed by Abbasi et al.~\cite{HiRes} did not confirm this result for the BL objects, but they found a significant correlation (revealed by Gorbunov et al.~\cite{Gorbi}) of cosmic rays with energies above $10^{19}$ eV and all events without selection in energy with high-polarization (HP) objects, as well as with the combined BL+HP sample from the same catalog. Abbasi et al.~\cite{HiRes} interpreted this correlation as an indication of the possible flux of neutral particles generated by Lacertae.

Other AGNs, namely Seifert galaxies closer than 40 Mpc, were treated as possible sources of UHECRs by Uryson~\cite{Uryson}. She found angular correlation of the showers of energies $3.2\times10^{19}<E<3\times10^{20}$ eV detected by the Akeno, AGASA, Haverah Park, and Yakutsk arrays with Seifert galaxies at redshifts $z\leq 0.0092$, which are faint x-ray and radio sources. Subsequently, correlation of UHECR arrival directions with these objects, as well as with other possible sources, was re-analyzed in~\cite{Compare}. It was shown that correlation with Seifert galaxies is observed only in the AGASA data, and if the deflection of particles in the Galactic magnetic field is taken into account, i.e., under the conditions different from those accepted in~\cite{Uryson}.

The Pierre Auger collaboration (PAC)~\cite{Auger} recently analyzed a sample consisting of 81 EASs with energies above $40$ EeV detected from January 1, 2004 to August 31, 2007. The authors used a part of the data (to May 27, 2006) in order to determine the parameters resulting in the maximum correlation of UHECR arrival directions with AGNs. Then, the second part of the data was used to confirm the hypothesis obtained.

\begin{figure}[t]
\center{\includegraphics[width=\columnwidth]{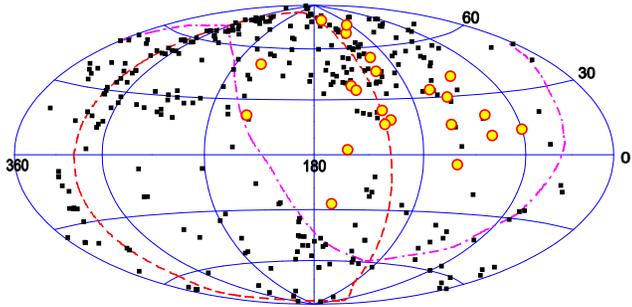}}
\caption{Fig.\ref{fig:Map}. Distribution of UHECR arrival directions (circles) and active nuclei (squares) in the Galactic coordinates. The dashed line is the Supergalactic plane and the dash-dotted line is bordering the observation region of the Yakutsk array.}
\label{fig:Map}\end{figure}

As a result, the observed UHECR arrival directions are found to be anisotropic, and there is a significant correlation of EASs with energies above $56$ EeV within an angle of $\psi=3.1^0$ with AGNs from catalog~\cite{VCV} located at distances $z\leq 0.018$ from the Earth (closer than 75 Mpc if the Hubble constant is equal to 71\,km\,s$^{-1}$Mpc$^{-1}$). In the second part of the data (from May 27, 2006), 8 of 13 EASs correlate with AGNs under the same conditions that have been found for the first part of the data, while the number of expected coincidences is 2.7 in the isotropic case. This corresponds to the chance probability $P=1.7\times10^{-3}$ for the uniform distribution.

In this work, in order to confirm or reject the PAC result assumed as the null hypothesis, an independent sample of $51$ EASs with energies above $40$ EeV and zenith angles below $60^0$ that were detected with the Yakutsk array and reported by Pravdin et al.~\cite{Pune} is analyzed. The error in the determination of the arrival angles of these showers is less than $5^0$. The energy of a primary particle initiating EAS is estimated on the basis of the total flux measurement of Cherenkov light and the numbers of electrons and muons at the observation level. The procedure used was described in~\cite{Mono,CRIS,JETP}. The error in the estimation of the energy, $\delta E/E$, is about $30\%$ and less than $50\%$ for the showers with axes inside the array area and in the effective region outside, respectively.

The Galactic coordinates of AGNs from~\cite{VCV} at the distances $z < 0.015$ are shown in Fig. \ref{fig:Map} in the Hammer-Aitoff projection along with the arrival angles of $22$ EASs with energies above $60$ EeV. The observation regions of the Pierre Auger observatory and Yakutsk array are directed towards and outwards the center of the Galaxy, respectively, because these two arrays are located in different hemispheres. Thus, these two observation regions are complementary and, in particular, cover different segments of the Supergalactic plane. The showers detected with the Yakutsk array seem to gather around the Supergalactic plane: the density of UHECRs is higher near this plane.

\begin{figure}[t]
\center{\includegraphics[width=0.75\columnwidth]{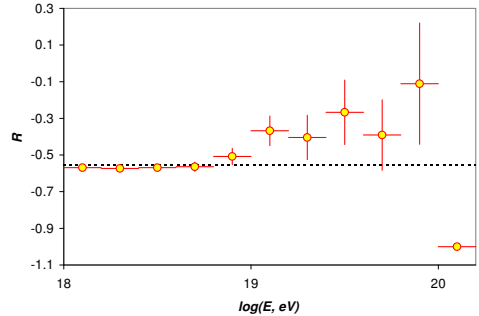}}
\caption{Fig.\ref{fig:sgpe}. Excess of UHECRs from the Supergalactic plane as a function of energy. The vertical bars are statistical errors and the horizontal bars are energy bins. The dashed line shows the value expected for the isotropic distribution.}
\label{fig:sgpe}\end{figure}

To verify this hypothesis by Stanev et al.~\cite{Stanev} using the extended Yakutsk array database, the excess $R=(n_{SGP}-n_{other})/(n_{SGP}+n_{other})$ of UHECRs arriving from the $\pm10^0$ vicinity of the Supergalactic plane, $n_{SGP}$, over the number from all other directions, $n_{other}$, is calculated and compared with the value $R_0=-0.555$ obtained for the isotropic distribution of cosmic rays taking into account different exposures of celestial regions over the array (see Fig. \ref{fig:sgpe}). An exposure was calculated using the algorithm described in~\cite{Wavelet}. As it is seen in the figure, $R$ tends to increase with energy, but there is no statistically significant excess of cosmic ray flux from the Supergalactic plane in the data in any energy bin. Glushkov and Pravdin~\cite{SGP} previously claimed that the observed UHECR flux from this plane noticeably exceeds the flux expected for the isotropic distribution if the angular interval width near the plane and energy threshold for the sample are adjusted. In particular, the excess in a narrow angular bin ($1^0-2^0$) and for the particle energies above $8$ EeV is $(4-5)\sigma$ for the Yakutsk array data. However, the correction factor to this excess that should be introduced in order to take into account a posteriori optimization of the data selection criteria is indefinite in this case.

\begin{figure}[t]
\center{\includegraphics[width=0.85\columnwidth]{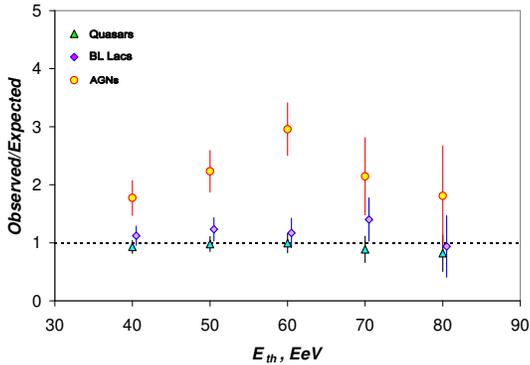}}
\caption{Fig.\ref{fig:others}. Ratio of the coincidences of UHECR arrival directions with extragalactic objects to the number of random coincidences expected for the isotropic distribution versus the threshold energy of the particles, $E>E_{th}$. The statistical error bars are also shown.}
\label{fig:others}\end{figure}

Let us proceed to the correlation of UHECRs detected in Yakutsk~\cite{Pune} with extragalactic objects. The test of the PAC hypothesis with the recommended parameters ($E>56$ EeV, $z<0.018$, and $\psi=3.1^0$) shows that the arrival directions of 12 of 24 EASs correlate with the AGNs from~\cite{VCV}, while the number of coincidences expected for the isotropic distribution is $5.6$. In this case, the chance probability of 12 or more coincidences is $P=4\times10^{-3}$. Therefore, the Yakutsk array data confirm the result obtained by the Pierre Auger collaboration, but at a lower significance level.

Due to the different observation region, as well as the energy/arrival angles estimation procedures, the 'optimal' parameters of correlation can be different for the Yakutsk array data. For this reason, scanning in the energy ($E>40$ EeV), redshift ($0.001<z<0.03$), and angular distance ($1^0<\psi<6^0$) is performed to determine the maximum ratio of the difference in the observed number of coincidences and the number expected for the isotropic case to the standard deviation. The maximum ratio appears to be reached for $22$ EASs with energies above $60$ EeV, $12$ of which arrive within $\psi=3^0$ of the AGNs (while an expected number is $4.1$) at distance from the Earth less than $z=0.015$ ($63$ Mpc). The chance probability is $P=2\times10^{-4}$. In this case, it is also necessary to determine a penalty factor to the probability due to a posteriori selection of the parameters; or the chosen parameters should be used as a null hypothesis for the testing with independent data.

Figure \ref{fig:others} shows the ratio of the observed number of coincidences of UHECR arrival directions (hereafter, within $3^0$) with quasars, Lacertae, and AGNs from~\cite{VCV} to the number of random coincidences expected for the isotropic distribution, taking into account the exposure of sky regions to the Yakutsk array. The active nuclei were chosen at the distances $z<0.015$, the quasars were taken at the distances $z<0.3$, and BL Lacs were selected with luminosities $m<18$ as in~\cite{TT}. According to Fig. \ref{fig:others}, there is no significant deviation of the number of coincidences from isotropic expectation in the case of Lacertae and quasars. Variation in the redshift ranges for quasars does not reveal any significant deviation. Any significant correlation is also absent for HP objects and BL+HP objects from~\cite{VCV}.

\begin{figure}[t]
\center{\includegraphics[width=0.8\columnwidth]{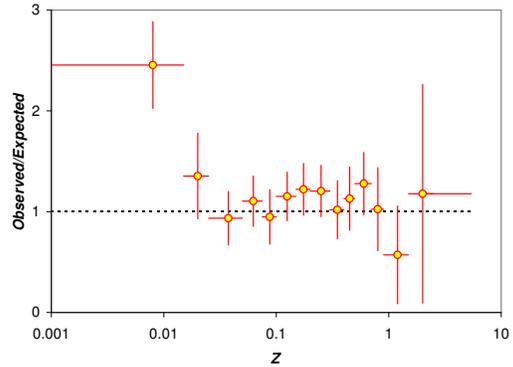}}
\caption{Fig.\ref{fig:z}. Ratio of the coincidences of UHECR arrival directions with AGNs to the expected number in various redshift bins, $z$. The vertical bars are statistical errors. The boundaries of $z$ (shown by the horizontal bars) are 0.001, 0.015, 0.025, 0.05, 0.075, 0.1, 0.15, 0.2, 0.3, 0.4, 0.5, 0.7, 0.9, 1.5, and 5.4.}
\label{fig:z}\end{figure}

Correlations with AGNs are found in the PAO and Yakutsk array data at almost the same angular distances ($3.1^0$ and $3^0$, respectively). It is unclear whether this is a random coincidence. The error in arrival angles of the showers with axes inside and outside the array area is estimated in~\cite{Pune} as $3^0$ and $5^0$, respectively. At the same time, as the energy increases, the angular error should decrease because the number of array stations fired in a shower increases with energy. Therefore, the actual accuracy of UHECR angles for the Yakutsk array can be equal to $3^0$ or better.

Approximate coincidence in the threshold energies at which the maximum correlation is observed by the two arrays may be accidental. A comparison shows the similarity in the shapes of the energy spectra measured by giant EAS arrays: if the correction factors are applied to the energy estimates of the primary particles inducing EASs, the spectra almost coincide both in shape and in intensity~\cite{Dip}. Systematic energy corrections introduced in this way for the Pierre Auger observatory and Yakutsk EAS array can differ from each other by a factor of $1.5-2$. Hence, the threshold energy at which the maximum correlation is reached for the PAO data would be from $90$ to $120$ EeV if the assumption about correction factors is valid.

Another part of the PAC hypothesis is that cosmic rays correlate with the directions to the AGNs at distances $z<0.018$ due to the Greisen-Zatsepin-Kuzmin effect~\cite{GZK}, which strongly suppress the flux of cosmic rays with energies $E>60$ EeV from cosmological distances. In order to verify this effect with the Yakutsk array data, the ratio of the observed number of coincidences of cosmic rays ($E>60$ EeV) with AGNs in various redshift bins to the number of random coincidences expected for the isotropic distribution is used (Fig. \ref{fig:z}). Indeed, a significant correlation of UHECRs with AGNs is found only in the interval $z\in(0.001,0.015)$. In all other redshift bins, the observed number of coincidences is equal (within errors) to the number expected in the isotropic case. Hence, this can be considered as one of the independent evidences of the Greisen-Zatsepin-Kuzmin effect.

It should be noted that the Yakutsk data, as in the case of the PAO data, possibly correlate not only with AGNs, but also with other astrophysical objects with a similar spatial distribution that are not considered in this work. Moreover, Gorbunov et al.~\cite{CenA} stated contrary to the PAC hypothesis: the conclusion that the most part of UHECRs are protons originating in nearby extragalactic sources (AGNs) can be rejected at 99\% CL. Instead, they explained the data~\cite{Auger} by the existence of a bright source in the direction of the Centaurus Supercluster. Another interpretation of these data was proposed by Wibig and Wolfendale~\cite{WW}, who stated that cosmic rays are nuclei with $<lnA>=2.2.\pm0.8$ that are generated in nearby (about tens of Mpc) radio galaxies.

Analysis of the muon detectors data of the Yakutsk array~\cite{Muons} shows that the data at energies above $10$ EeV can be explained within the framework of the two-component model of the cosmic ray composition consisting of protons and a considerable fraction of heavy nuclei. This is hardly consistent with the correlation of the arrival directions of such particles within $3^0$ with AGNs. 'Clusters' of EASs associated with clusters of (radio) galaxies are hardly seen in the data of the Yakutsk array in contrast to the PAO data.

\begin{figure}[t]
\center{\includegraphics[width=\columnwidth]{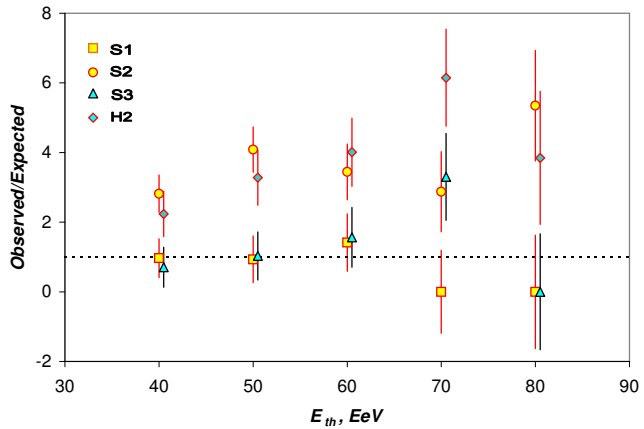}}
\caption{Fig.\ref{fig:Seyferts}. Ratio of UHECR coincidences with different classes of AGNs to the expected number.}
\label{fig:Seyferts}\end{figure}

If the hypothesis that active galactic nuclei are sources of UHECRs is accepted, the following question arises: Whether all classes of the objects belonging to active galaxies are sources or, for example, only Seifert galaxies generate cosmic rays, as was assumed by Uryson~\cite{Uryson}. To answer this question, one can analyze a correlation of UHECR arrival directions with each class of the objects separately. Figure~\ref{fig:Seyferts} shows the results of such an analysis of the data. Here, AGNs are divided into four classes according to the proposal of V\'{e}ron­Cetty and V\'{e}ron~\cite{VCV}: i) S1, Seifert galaxies of the first type with broad Balmer lines; ii) S2, Seifert galaxies of the second type; iii) S3, so-called LINERs (low-ionization nuclear emission-line regions), which are galaxies with weak nuclear emission lines; and iv) H2, galaxies whose spectrum of nuclear emission lines is similar to that of nebulae ionized by hot stars. The redshift boundaries giving the maximum correlation with the Yakutsk data are selected for each class of the objects: $z<0.015$ for S1 and S3, $z<0.016$ for S2, and $z<0.024$ for H2.

Correlations of the Seifert galaxies of the first type and LINERs with cosmic rays do not exceed correlation expected for the isotropic distribution. Only S2 and H2 objects correlate with UHECRs (with the maxima at $E>50$ and $E>70$ EeV, respectively). The excess of the observed number of coincidences over the number expected for the isotropic distribution is $4.7\sigma$ and $3.7\sigma$ in the maxima for S2 and H2, respectively. Therefore, possible sources of UHECRs are Seifert galaxies of the second type and/or H2 objects at distances less than $100$ Mpc.

This work is supported by the Russian Foundation for Basic Research (project no. 06-02-16973) and by the Council of the President of the Russian Federation for Support of Young Scientists and Leading Scientific Schools (project no. NSh-7514.2006.2).

\vspace{0.5cm}
{\sl Translated by R. Tyapaev}.


\vspace{0.5cm}

\begin{thebibliography}{15}
\bibitem{Farrar} G.\,R. Farrar and P.\,L. Biermann, Phys. Rev. Lett. {\bf 81}, 3579 (1998).

\bibitem{Glushkov} A. V. Glushkov, Phys. At. Nucl. {\bf 68}, 237 (2005).

\bibitem{Waxman} E. Waxman, Phys. Rev. Lett. {\bf 75}, 386 (1995);
M. Vietri, Astrophys. J. {\bf 453}, 883 (1995).

\bibitem{GRB} L. Anchordoqui for the P. Auger collaboration, Proc. 30$^{th}$ ICRC, Merida (2007) icrc1041;
arXiv:astroph/0706.0989.

\bibitem{Stanev} T. Stanev, P. Biermann, J. Lloyd-Evans et al., Phys. Rev. Lett. {\bf 75}, 3056 (1995).

\bibitem{Uchihori} Y. Uchihori et al., Astropart. Phys. {\bf 13}, 151 (2000).

\bibitem{TT} P.\,G. Tinyakov and I.\,I. Tkachev, JETP Lett. {\bf 74}, 445 (2001);
P.\,G. Tinyakov and I.\,I. Tkachev, Astropart. Phys. {\bf 18}, 165 (2002);
D.\,S. Gorbunov, P.\,G. Tinyakov, I.\,I. Tkachev, S.\,V. Troitsky, Astrophys. Journ. {\bf 577}, L93 (2002).

\bibitem{VCV} M.\,-P. V\'{e}ron-Cetty and P. V\'{e}ron, Astron. Astrophys. {\bf 374},  92 (2001) - 10$^{th}$ edition;
M.\,-P. V\'{e}ron-Cetty and P. V\'{e}ron, Astron. Astrophys. {\bf 455}, 773 (2006) - 12$^{th}$ edition.

\bibitem{HiRes} R.\,U. Abbasi et al., Astrophys. Journ. {\bf 636}, 680 (2006).

\bibitem{Gorbi} D.\,S. Gorbunov, P.\,G. Tinyakov, I.\,I. Tkachev, S.\,V. Troitsky, JETP Lett. {\bf 80}, 145 (2004).

\bibitem{Uryson} A. V. Uryson, JETP Lett. {\bf 64}, 77 (1996);
A. V. Uryson, JETP {\bf 86}, 213 (1998);
A. V. Uryson, JETP {\bf 89}, 597 (1999).

\bibitem{Compare} D.\,S. Gorbunov, S.\,V. Troitsky, Astropart. Phys. {\bf 23}, 175 (2005);
arXiv:astro-ph/0410741.

\bibitem{Auger} The Pierre Auger Collaboration, Science {\bf 318}, 938 (2007).

\bibitem{Pune} M.\,I. Pravdin, A.\,V. Glushkov, A.\,A. Ivanov et al., Proc. 29$^{th}$ ICRC, Pune, {\bf 7}, 243 (2005).

\bibitem{Mono} M. N. D'yakonov, T. A. Egorov, N. N. Efimov, et al.,  {\sl Cosmic Radiation of Extremely High Energy}, (Nauka, Novosibirsk, 1991) [in Russian].

\bibitem{CRIS} V.\,P. Egorova, A.\,V. Glushkov, A.\,A. Ivanov et al., Nucl. Phys. B (Proc. Suppl.), {\bf 136}, 3 (2004).

\bibitem{JETP} S. P. Knurenko, A. A. Ivanov, I. E. Sleptsov, and A. V. Saburov, JETP Lett. {\bf 83}, 473 (2006);
S. P. Knurenko, A. A. Ivanov, and A. V. Saburov, JETP Lett. {\bf 86}, 621 (2007);
A. A. Ivanov, S. P. Knurenko, and I. E. Sleptsov, JETP {\bf 104}, 872 (2007).

\bibitem{Wavelet} A. A. Ivanov, A. D. Krasil'nikov, and M. I. Pravdin, JETP Lett. {\bf 78}, 695 (2003);
A. A. Ivanov, Thesis, University of Yakutsk, 2005.

\bibitem{SGP} A. V. Glushkov and M. I. Pravdin, JETP {\bf 92}, 887 (2001);
A. V. Glushkov and M. I. Pravdin, Astron. Lett. {\bf 27}, 493 (2001).

\bibitem{Dip} A. A. Ivanov, Dokl. Phys. {\bf 52}, 523 (2007);
V. Berezinsky, Proc. 30$^{th}$ ICRC, Merida (2007).

\bibitem{GZK} G. T. Zatsepin and V. A. Kuz'min, Sov. Phys. JETP Lett. {\bf 4}, 53 (1966);
K. Greisen, Phys. Rev. Lett. {\bf 16}, 748 (1966).

\bibitem{CenA} D. Gorbunov, P. Tinyakov, I. Tkachev, S. Troitsky, arXiv:astroph/0711.4060.

\bibitem{WW} T. Wibig, A.\,W. Wolfendale, arXiv:astroph/0712.3403.

\bibitem{Muons} A.\,V. Glushkov, D.\,S. Gorbunov, I.\,T. Makarov et al., JETP Lett. {\bf 87}, 220 (2008);
arXiv:astroph/0710.5508.

\end{thebibliography}
\end{document}